\documentclass[useAMS,usegraphicx,usenatbib]{mn2e}
\title[New Method for Exploring Super-Eddington AGNs by NIR]
{New Method for Exploring Super-Eddington AGNs by 
Near-infrared Observations}
\author[N. Kawakatu et al.]{N. Kawakatu$^{1}$
\thanks{E-mail:kawakatu@ccs.tsukuba.ac.jp (NK)} 
and K. Ohsuga $^{2}$ \\
$^{1}$ Graduate School of Pure and Applied Sciences, 
University of Tsukuba, 1-1-1 Tennodai, Tsukuba 305-8571 \\
$^{2}$ National Astronomical Observatory of Japan, 2-21-1 Osawa, 
Mitaka, Tokyo 181-8588, Japan}
\begin{document}

\date{}

\pagerange{\pageref{firstpage}--\pageref{lastpage}} \pubyear{2011}
\twocolumn

\maketitle

\label{firstpage}

\begin{abstract}
We propose a new method to explore the candidate super-Eddington 
active galactic nuclei (AGNs). We examine the properties of infrared (IR) 
emission from the inner edge of the dusty torus in AGNs,
which are powered by super- or sub-Eddington accretion flows around 
black holes, by considering the dependence of the polar angle on the 
radiation flux of accretion flows \citep{Wa05}. 
We find that for super-Eddington AGNs, of which the mass accretion rate 
is more than $10^{2}$ times larger than the Eddington rate, the ratio 
of the AGN IR luminosity and the disc bolometric luminosity is less than 
$10^{-2}$, unless the half opening angle of the torus ($\theta_{\rm torus}$) 
is small ($\theta_{\rm torus}<65^\circ$). 
This is due to the self-occultation effect, whereby the self-absorption 
at the outer region of the super-Eddington flow dilutes the illumination 
of the torus. Such a small luminosity ratio is not observed in sub-Eddington 
AGNs, whose mass accretion rate is comparable to or no more than 10 times larger than the Eddington mass accretion rate, except for extremely thin tori 
($\theta_{\rm torus}>85^\circ$). 
We also consider the properties of the near-IR (NIR) emission 
radiated from hot dust $>$ 1000 K. We find that super-Eddington 
AGNs have a ratio of the NIR luminosity 
to the bolometric luminosity, $L_{\rm NIR,AGN}/L_{\rm bol,disc}$, 
at least one order of magnitude smaller than 
for sub-Eddington AGNs for a wide range of half opening angle 
($\theta_{\rm torus} > 65^{\circ}$), for various 
types of dusty torus model. Thus, a relatively low $L_{\rm NIR,AGN}
/L_{\rm bol,disc}$ is a property that allows identification of candidate 
super-Eddington AGNs. Lastly, we discuss the possibility that NIR-faint 
quasars at redshift $z\sim 6$  discovered by a recent deep SDSS survey 
may be young quasars whose black holes grow via super-Eddington accretion.

\end{abstract}

\section{Introduction}
It is now widely accepted that most normal galaxies have central 
supermassive black holes (SMBHs) with $M_{\rm BH} =10^{6\rm -9}M_{\odot}$ 
(e.g.,\citealt{KR95,Mi95,Gi09}).
High resolution observations of nearby galaxies have revealed a close 
correlation between the mass of SMBHs and the bulge mass, or the velocity 
dispersion of the bulge (e.g., \citealt{KR95,FM00,MH03,HR04}). 
This implies that the formation of SMBHs is strongly coupled with 
the formation of galaxies. However, the formation and evolution of SMBHs is 
still an open question and a hot topic in astrophysics. 

The discovery of luminous quasars at redshift $z >6$  shows that SMBHs of the 
order of billions of solar masses exist at the end of the re-ionization epoch 
(e.g., \citealt{Fa01,Fa06,Go06,Wi07,Wi10}). Considering Eddington-limited 
accretion, $z > 6$ quasars may not form readily. Unless the seed black hole 
mass is large, gas accretion with low radiative efficiency and/or BH mergers 
significantly contribute to the growth of SMBHs (see \citealt{Sh05}). 
If super-Eddington accretion is possible, the growth timescale of BHs can be 
much shorter than the Eddington timescale (e.g., \citealt{Ka04,Oh05}). Thus, 
super-Eddington accretion might explain the formation of SMBHs. 
Recently, \citet{KW09} predicted that super-Eddington accretion is 
required for the formation of $z>6$ quasars with $M_{\rm BH}\approx 
10^{9}M_{\odot}$, because the final BH mass is greatly suppressed by star 
formation in a circumnuclear disc and AGN outflow based on the coevolution 
model of SMBH growth and a circumnuclear disc (\citealt{KW08}; see also 
\citealt{VR05}).

The occurrence of super-Eddington accretion flow has not yet been accurately 
verified, although this issue has been investigated since the 1970s (e.g., 
\citealt{SS73,Be78,Ab88,HC91,WF99,Wa00}. By using two-dimensional radiation 
hydrodynamic simulations, \citet{Oh05} demonstrated for the fist time that 
quasi-steady super-Eddington disc accretion is possible, because both the 
radiation field and the mass accretion flow around a BH are non-spherical 
and the large number of photons generated in the disc is swallowed by 
the BH without being radiated away due to photon-trapping (see \citealt{OM07}). This conclusion has recently been confirmed by two-dimensional 
radiation magnetohydrodynamic simulations \citep{Oh09}. However, it is 
still unclear whether super-Eddington accretion can significantly contribute to the formation of SMBHs, because the radiation pressure and/or disc wind 
associated with super-Eddington accretion might play an important role in the gas accretion process in host galaxies 
(e.g., \citealt{SR98,Fa99,Ki03,Oh07}).

In the local Universe, we have discovered a class of AGNs with a higher 
Eddington ratio among the AGN population, i.e., Narrow Line Seyfert 1 galaxies 
(NLS1s). They have characteristic properties such as narrow Balmer lines 
(e.g., \citealt{OP85,Po00}), strong soft X-ray excess 
(e.g., \citealt{Po95,Bo96,Bo03}) and rapid variability (e.g., \citealt{Ot96,Le99,Bo02,Ga04}). These properties indicate that they have a small black 
hole mass and high Eddington ratio (e.g., \citealt{BB98,Ha00,Mi00}). 
The observed bolometric luminosity of NLS1s saturates at a few times 
the Eddington luminosity (e.g., \citealt{CK04,Co06,Mu08,La09}), which is 
consistent with predictions of the super-Eddington accretion disc 
(e.g., \citealt{Wa00,Oh05,Oh09}). In addition, the star formation rate 
of NLS1 hosts is higher than that of BLS1s (\citealt{Sa10}). 
Thus, NLS1s may be the early phase of rapid BH growth via super-Eddington 
accretion (e.g., \citealt{Ma00,Ka04}). In the high-$z$ universe, we may speculate that super-Eddington AGNs are a more dominant population of AGNs because the lifetime of SMBH growth is closer to the cosmic age if the formation of SMBHs is mainly a gas accretion process. To confirm this, it is important to determine whether the fraction of super-Eddington AGNs increases with redshift. Some observations to date have indicated that the average Eddington ratio increases  slightly with redshift (e.g., \citealt{Ko06,Sh08,Wi10}). However, we have not observed super-Eddington AGNs among high-$z$ quasars (e.g., \citealt{MD04,Sh08}) and AGNs in high-$z$ massive galaxies (e.g., \citealt{Ya09}). Thus, it is essential to find candidate super-Eddington AGNs before starting detailed observations.

According to unification models of AGNs, the accretion disc is surrounded by a 
dusty structure (e.g., \citealt{An93,El08}). A significant fraction of the emitted ultraviolet and optical radiation of the accretion disc is absorbed by the dust and is re-emitted at IR wavelengths. In particular, the hot dust is 
directly heated by the central engine and produces near-IR (NIR) emission (e.g., \citealt{Ri78,Ha03}). Since it has been reported that the polar angle 
dependence of the radiation flux for super-Eddington accretion flow deviates from that for sub-Eddington flow (e.g., \citealt{Fu00,Oh05,Wa05} 
hereafter W05), the reprocessed IR emission may be useful for exploring 
candidates of super-Eddington accretion flow in AGNs. In this paper, we investigate the properties of IR (as well as NIR) emission from the inner edge of the 
dusty torus, employing radiative flux from the disc accretion flows by W05, 
in which the radiation fluxes of sub- and super-Eddington accretion discs are 
given as functions of the polar angle and the mass accretion rate. 

\section{Radiation Flux from Accretion disc}
We briefly summarize the relation between the radiation flux 
from the accretion disc and the mass accretion rate into an SMBH. 
For a small mass accretion rate, i.e., $\dot{m}_{\rm BH}\equiv
\dot{M}_{\rm BH}/(L_{\rm Edd}/c^2)=1$--$10$, where $\dot{M}_{\rm BH}$ 
and $L_{\rm Edd}$ are the mass accretion rate into an SMBH and the 
Eddington luminosity, respectively, if the accretion disc is geometrically thin, 
it is known as a standard disc \citep{SS73}. 
As the mass accretion rate increases ($\dot{m}_{\rm BH}\geq 10$), 
the disc becomes geometrically thick via strong radiation pressure, 
i.e., a slim disc, as first introduced by \citet{Ab88}. 
The maximum thickness of the disc is about $45^{\circ}$ because the ratio 
of the scale height and the disc size is approximately unity (e.g., \citealt{Ka08}. 
W05 examined the observed spectra for various mass accretion rates 
and included the effect of disc geometry. 
They found that the radiation flux is proportional to $\cos\theta$ for 
the sub-Eddington regime, $\dot{m}_{\rm BH}=1$--$10$, where $\theta$ is the 
polar angle (see Fig. 1). The factor $\cos{\theta}$ represents the change of 
the effective area via a change of $\theta$, i.e., the projection effect. 
For a super-Eddington accretion disc with $\dot{m}_{\rm BH} \geq 10^2$,
although the radiation flux is also proportional to $\cos\theta$ 
at $\theta < 45^{\circ}$, it decreases more significantly as $\theta$ increases
in regions with $\theta \geq 45^{\circ}$.

The bolometric luminosity of an accretion disc is expressed as a function of 
$\dot{m}_{\rm BH}$ \citep{Wa00} as follows: 
\begin{equation}
L_{\rm bol, disc}=\left \{
 \begin{array}{l}
 2\left(1+\ln{\frac{\dot{m}_{\rm BH}}{20}}\right)
L_{\rm Edd}\,\,\, (\dot{m}_{\rm BH} \geq 20), \\ \\
 \left(\frac{\dot{m}_{\rm BH}}{10}\right)
L_{\rm Edd} \,\,\,\,\, (\dot{m}_{\rm BH} < 20).
 \end{array}\right .
\end{equation}

Based on the results of W05, 
we can describe the radiation flux from the accretion disc, 
$F(\theta)$, 
as follows:

For $\theta < 45^{\circ}$, 
\begin{equation}
F(\theta)=\frac{L_{\rm bol,disc}}{4\pi r^{2}} \left \{
 \begin{array}{l}
2\cos{\theta}\,\,\, (\dot{m}_{\rm BH}=1,10), \\ \\
\frac{20}{7}\cos{\theta}\,\,\, (\dot{m}_{\rm BH}=10^{2}), \\ \\
\frac{10}{3}\cos{\theta}\,\,\, (\dot{m}_{\rm BH}=10^{3}). 
 \end{array}\right.
\end{equation}

For $\theta \geq 45^{\circ}$,
\begin{equation}
F(\theta)=\frac{L_{\rm bol,disc}}{4\pi r^{2}} \left \{
 \begin{array}{l}
2\cos{\theta}\,\,\, (\dot{m}_{\rm BH}=1,10), \\ \\
\frac{40\sqrt{2}}{7}\cos^{4}{\theta}\,\,\, (\dot{m}_{\rm BH}=10^{2}), \\ \\
\frac{160}{3}\cos^{9}{\theta}\,\,\, (\dot{m}_{\rm BH}=10^{3}),
\end{array}\right.
\end{equation}
where $r$ is the distance from the central BH. 
The coefficients of the cosine functions (see eqs. (2) and (3)) 
are determined by setting the integral of $F(\theta)$ to be equal 
to $L_{\rm bol, disc}$ 
(see eq. (1)) and the continuity of the radiation flux $F(\theta)$ 
at $\theta=45^{\circ}$. 
We should note that eqs. (2) and (3) well reproduce the results 
of W05 (see Fig. 3 in W05). 

Note that we do not take the corona into account in the present
study(but see $\S 4.2$). Although X-ray observations have shown 
the radiation from the disc to be comptonized at the corona above 
the disc, the size and geometries of the corona are still 
open questions (e.g., \citealt{KD04}). 
The optical thickness of the corona is thought to be
around unity or less in the case of sub-Eddington discs 
(e.g., \citealt{Kaw08}). Hence, the $\theta$-dependence of the radiative 
flux should not change much. 
Global radiation hydrodynamic simulations of super-Eddington flows
have revealed that the radiative flux is highly collimated, although 
hot rarefied plasma (corona) appears above the disc \citep{Oh05}.
Recently, similar results have been found by global radiation
magnetohydrodynamic simulations \citep{Oh09}. Thus, we assume
that the radiative flux of the sub-Eddington disc is proportional to 
$\cos\theta$ and employ the collimated radiative flux from the
super-Eddington disc in the present study. 

\begin{figure}
\begin{center}
\includegraphics[height=4.5cm,clip]{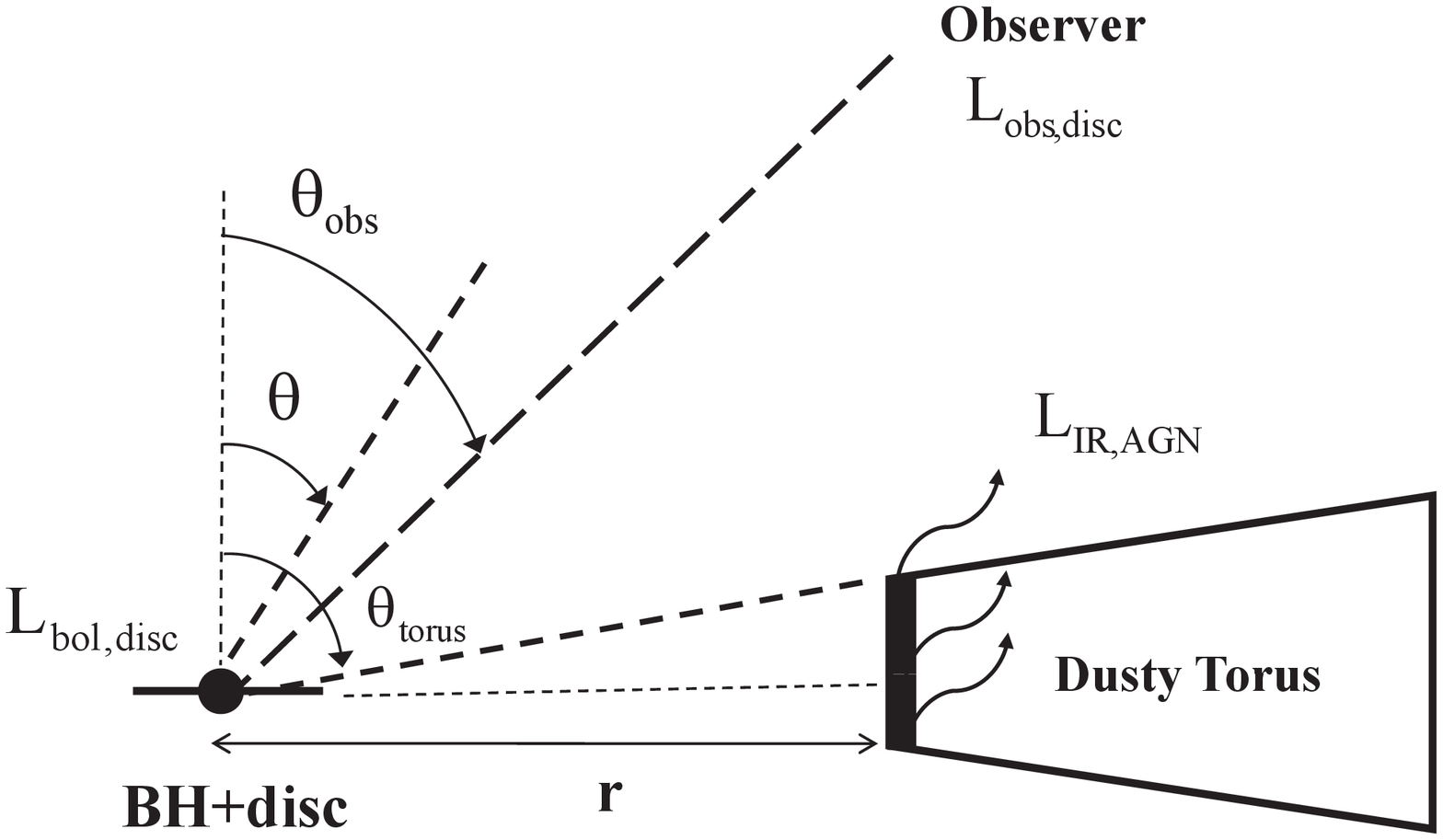}
\end{center}
\caption{
Model for a dusty torus around an SMBH plus an accretion disc. 
$L_{\rm bol,disc}$ is the bolometric luminosity of the accretion disc 
and $\theta$ is the polar angle. 
$\theta_{\rm torus}$, $r$ and 
$L_{\rm IR, AGN}$ are the half opening angle of the torus, the distance 
from the BH and the infrared luminosity re-emitted by the inner edge of 
the dusty torus, respectively. $\theta_{\rm obs}$ is the 
inclination angle.
}
\end{figure}

\section{Signature of Super-Eddington Accretion Flows in AGNs}
\subsection{Infrared luminosity}
In order to examine the AGN IR properties for various mass 
accretion rates, we model a dusty torus around an SMBH 
plus an accretion disc system (Fig. 1). 
Assuming that the radiation of the accretion disc is absorbed by 
the dust and the AGN IR radiation is re-emitted isotropically 
at the inner edge of the torus, the AGN IR luminosity ($L_{\rm IR, AGN}$) 
depends on only the half opening angle of the torus, $\theta_{\rm torus}$, 
and is independent of the structure of the inner edge of the torus. 
In other words, energy conservation holds at the inner edge of the torus. 
Then, $L_{\rm IR, AGN}(\theta_{\rm torus})$ can be given by 
\begin{equation}
L_{\rm IR, AGN}(\theta_{\rm torus})=4\pi r^{2}
\int_{\theta_{\rm torus}}^{\pi/2}F(\theta)\sin{\theta}d\theta.  \\
\end{equation}
where $r$ is the distance from the central BH. 
Note that eq. (4) is valid even in a clumpy model (e.g., \citealt{Ne02,Ne08, 
WN02,Ho06,Mor09}) 
since there are several optically thick clouds along the line of 
sight and thus all the radiation of 
the accretion disc is absorbed by clouds near a BH . 

Substituting eqs. (2) and (3) into eq. (4), the ratio of 
the AGN IR luminosity and the bolometric luminosity is obtained as follows: 

For $\theta_{\rm torus} < 45^{\circ}$,
\begin{equation}
\frac{L_{\rm IR,AGN}}{L_{\rm bol,disc}}=\left \{
\begin{array}{l}
\cos^{2}{\theta_{\rm torus}},\,\,\, (\dot{m}_{\rm BH}=1,10), \\ \\
\frac{10}{7}\left(\cos^{2}{\theta_{\rm torus}}-\frac{3}{10} \right), \,\,\, 
(\dot{m}_{\rm BH}=10^{2}), \\ \\
\frac{5}{3}\left(\cos^{2}{\theta_{\rm torus}}-\frac{2}{5} \right),\,\,\, 
(\dot{m}_{\rm BH}=10^{3}).
 \end{array}\right.
\end{equation}

For $\theta_{\rm torus} \geq 45^{\circ}$, 
\begin{equation}
\frac{L_{\rm IR,AGN}}{L_{\rm bol,disc}}=\left \{
\begin{array}{l}
\cos^{2}{\theta_{\rm torus}},\,\,\, (\dot{m}_{\rm BH}=1,10), \\ \\
\frac{8\sqrt{2}}{7}\cos^{5}{\theta_{\rm torus}}, \,\,\, (\dot{m}_{\rm BH}=10^{2}), \\ \\
\frac{16}{3}\cos^{10}{\theta_{\rm torus}},\,\,\, (\dot{m}_{\rm BH}=10^{3}).
 \end{array}\right.
\end{equation}
These relations are understood in terms of the projection effect, 
self-occultation and the covering factor of the dusty torus. 
For $\dot{m}_{\rm BH}=1$--$10$, the self-occultation effect does not occur 
because the disc is geometrically thin. Since the radiation flux is reduced 
by the projection effect, 
$F(\theta)\propto\cos\theta$ (see eqs. (2) and (3)), and since the covering 
factor of the torus is given by $4\pi \cos\theta_{\rm torus}$, the IR luminosity decreases with an increase of $\theta_{\rm torus}$. The radiation flux is more sensitive to the polar angle in super-Eddington accretion discs with $\dot{m}_
{\rm BH}\geq 10^2$ than in sub-Eddington accretion discs via the self-
occultation effect (see eqs. (2) and (3)). Hence, the IR luminosity for $\dot{m}
_{\rm BH}=10^2$ and $10^3$ drastically decrease as the half opening angle 
increases when $\theta_{\rm torus}\geq 45^{\circ}$ (see eq. (6)). 

\begin{figure}
\begin{center}
\includegraphics[height=6.5cm,clip]{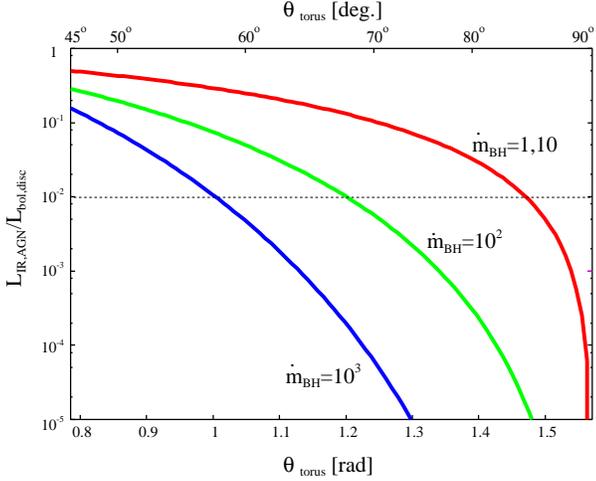}
\end{center}
\caption{
$L_{\rm IR,AGN}/L_{\rm bol,disc}$ against $\theta_{\rm torus}$ 
for various $\dot{m}_{\rm BH}$. 
The red line corresponds to $\dot{m}_{\rm BH}=1,10$, the green 
line shows the case for $\dot{m}_{\rm BH}=10^{2}$ and the blue 
line represents the case for $\dot{m}_{\rm BH}=10^{3}$.
}
\end{figure}

In Fig. 2, we plot the ratio of the IR luminosity to the bolometric luminosity, 
$L_{\rm IR,AGN}/L_{\rm bol,disc}$, against $\theta_{\rm torus}$ for various 
$\dot{m}_{\rm BH}$. As mentioned above, $L_{\rm IR,AGN}/L_{\rm bol,disc}$ decreases with $\theta_{\rm torus}$ for any $\dot{m}_{\rm BH}$, 
and this trend is much stronger as $\dot{m}_{\rm BH}$ becomes larger. 
Super-Eddington AGNs with $\dot{m}_{\rm BH} \geq 10^2$ 
have much smaller ratios of the IR and bolometric luminosities 
than sub-Eddington AGNs. 
Although the ratios of $L_{\rm IR,AGN}/L_{\rm bol,disc}$ for $\dot{m}_{\rm BH}
=10^2$ and $10^3$ are only a few times smaller than that for $\dot{m}_{\rm BH}=
1$--$10$ in the case of $\theta_{\rm torus}\approx 45^{\circ}$, they are one or two orders of magnitude lower than that for $\dot{m}_{\rm BH}\leq 10$ in 
a wide range of half opening angle, $\theta_{\rm torus} > 65^{\circ}$.
We also find that it is hard for sub-Eddington AGNs
to achieve a low $L_{\rm IR,AGN}/L_{\rm bol,disc}$, e.g., 
$L_{\rm IR,AGN}/L_{\rm bol,disc}< 10^{-2}$, 
unless the obscuring structure is extremely geometrically thin, 
e.g., $\theta_{\rm torus} > 85^{\circ}$. 
In contrast, 
we find $L_{\rm IR,AGN}/L_{\rm bol,disc}< 10^{-2}$
when $\theta_{\rm torus}>65^\circ$ for $\dot{m}_{\rm BH}=10^2$
and when $\theta_{\rm torus}>55^\circ$ for $\dot{m}_{\rm BH}=10^3$.
Thus, except for an extremely thin torus,
a low $L_{\rm IR,AGN}/L_{\rm bol,disc} (<10^{-2})$
is a signature of the occurrence of super-Eddington accretion flow 
with $\dot{m}_{\rm BH} \geq 10^2$.

These results are summarized as a schematic diagram in Fig. 3. 
The bottom left panel represents sub-Eddington accretion with 
a thick torus. 
In this case, $L_{\rm IR,AGN}/L_{\rm bol,disc}$ is maximal. 
For super-Eddington accretion ($\dot{m}_{\rm BH} \geq 10^2$) 
with a thick torus (top left panel), 
$L_{\rm IR,AGN}/L_{\rm bol,disc}$ is slightly smaller than 
the bottom left panel due to the self-occultation effect.
For sub-Eddington accretion with a thin torus 
(bottom right panel), 
$L_{\rm IR,AGN}/L_{\rm bol,disc}$ becomes smaller than the case 
of the bottom left panel, 
because the radiation of the accretion disc 
is reprocessed at the small inner surface of the torus.
Super-Eddington accretion ($\dot{m}_{\rm BH} \geq 10^2$) with a thin torus 
(top right panel) has the smallest $L_{\rm IR,AGN}/L_{\rm bol,disc}$ 
in the four panels of Fig. 3 due to the self-occultation effect of the disc
along with the small covering factor of the torus. 

\begin{figure}
\begin{center}
\includegraphics[height=6.5cm,clip]{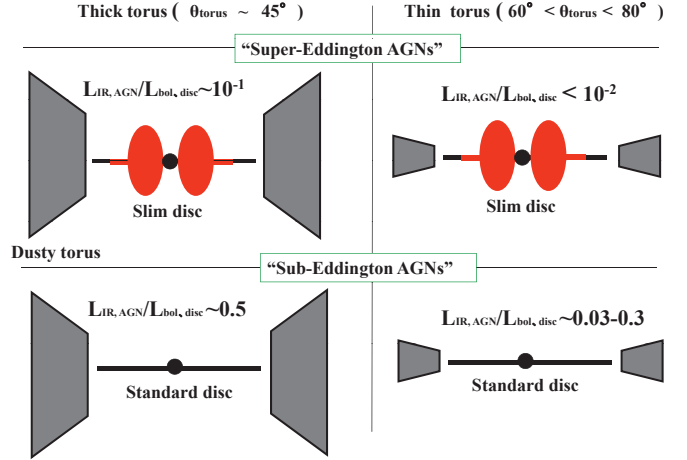}
\end{center}
\caption{
Summary of the dependence of $L_{\rm IR,AGN}/L_{\rm bol,disc}$ on 
the mass accretion rate into an SMBH and the geometry of the dusty torus. 
Typical values of $L_{\rm IR,AGN}/L_{\rm bol,disc}$ (see Fig. 2) 
are shown in each panel. 
 ``Super-Eddington AGNs" 
are expressed as AGNs with $\dot{m}_{\rm BH}\geq 10^{2}$ 
and ``sub-Eddington AGNs" are expressed as AGNs with $\dot{m}_{\rm BH}
=1,10$. 
}
\end{figure}

We here mention the effect of the inclination angle, $\theta_{\rm obs}$ 
(Fig. 1). The observed ratio $L_{\rm IR,AGN}/L_{\rm obs,disc}
(\theta_{\rm obs})$ depends on $\theta_{\rm obs}$, but $L_{\rm IR,AGN}$ 
is independent of $\theta_{\rm obs}$, where $L_{\rm obs,disc}(\theta_
{\rm obs})$ is the observed disc luminosity at the viewing angle of 
$\theta_{\rm obs}$. 
However, the difference between $\theta_{\rm obs}=0^{\circ}$ (face on) and 
$\theta_{\rm obs}$=30$^{\circ}$ is a factor of $\sqrt{3}/2$. 
In contrast, in the case of 45 deg $<\theta_{\rm obs} < \theta_{\rm torus}$,
the dependence of $\theta_{\rm obs}$ on $F(\theta_{\rm obs})$ is 
sensitive to $\dot{m}_{\rm BH}$. For example, the difference 
between $\theta_{\rm obs}$=0$^{\circ}$ (face on) and $\theta_{\rm obs}$=60
$^{\circ}$ is a factor of $1/2$, $\sqrt{2}/8$ and $1/32$ for $\dot{m}_{\rm BH}=1$--$10$, $10^{2}$ and $10^{3}$, respectively. Thus, when $\theta_{\rm obs} > 
45$ deg, the IR luminosity of super-Eddington AGNs could be similar to 
that of sub-Eddington AGNs. 

In summary, the IR luminosity of super-Eddington AGNs is very faint for $\theta_{\rm obs} < 45^{\circ}$, while it is not so faint for 45$^{\circ}<\theta_{\rm obs} < \theta_{\rm torus}$. Therefore, in order to examine the candidate super-Eddington AGNs effectively by IR observations, it is worth constructing  samples of pure type 1 AGNs (i.e., $\theta_{\rm obs} < 45^{\circ}$) rather than 
type 1.5 and 1.8 AGNs.

In this paper, we assume that the dusty torus is aligned with the 
accretion disc, such as in Fig. 1. However, this is not a trivial assumption because it is unclear whether the dusty torus is a reservoir of the gas which 
accretes onto the accretion disc. If the accretion disc is misaligned 
with the dusty torus, the self-occultation effect becomes less important 
as the angle between the rotation axis of the accretion disc and that of the torus increases (see eqs. (2) and (3)), since a large number of photons from the 
accretion disc are absorbed and re-emitted at the dusty torus. Thus, the ratio 
$L_{\rm IR,AGN}/L_{\rm bol,disc}$ for super-Eddington AGNs would be 
comparable to that for sub-Eddington AGNs. 

Lastly, we note the relation between the AGN type and 
$\theta_{\rm torus}$. It may be questioned whether our method is worthwhile 
if the AGN types are mainly determined by $\theta_{\rm torus}$ and 
$\theta_{\rm torus}=45^{\circ}$ is a standard torus model for all types of AGN. However, recent observations on circum-nuclear regions of AGNs have revealed 
that type 2 Seyferts are more frequently associated with starbursts than 
type 1 Seyferts (e.g., \citealt{He89,Ma97,PC96,Hu97,Ma98,Sc99,St00,Go01}). 
In addition, the fraction of type 2 AGNs decreases with AGN luminosity 
(e.g., \citealt{Ue03,Ma07,Has08}). These findings imply that the type1/type2 ratio is not simply determined by the opening angle ($\theta_{\rm torus}$). 
Moreover, the formation mechanism of the obscuring torus is still a hotly debated issue theoretically (e.g., \citealt{KB88,PK92,OU99,OU01,WN02,WU05,Sc09}). 
Thus, even if the effect we proposed becomes significant only for relatively 
large $\theta_{\rm torus}$ ($>65^{\circ}$), our method is useful for exploring the candidates of super-Eddington AGNs.

\subsection{Near-infrared luminosity}
Near-IR (NIR) emission is produced by the hot dust ($> 1000\,{\rm K}$) 
which is directly heated by the central AGNs, 
though the total IR emission may be contaminated by the nuclear starburst. 
Thus, based on the results of Fig. 2, we investigate the 
dependence of the ratio $L_{\rm NIR,AGN}/L_{\rm bol,disc}$ on 
the mass accretion rate, $\dot{m}_{\rm BH}$. 
The NIR luminosity, $L_{\rm NIR,AGN}$, strongly depends on 
the shape of the inner edge of the torus, because 
the hot dust temperature, $T_{\rm d}(\theta)$, is sensitive 
to the structure of the torus. We here consider the following 
three typical models for the structure of the inner edge of 
the torus (see Fig. 4). 

First, we examine the case where the inner radius of the torus $r_{\rm in}$ 
is determined by the sublimation radius $r_{\rm sub}(\theta)$ (Fig. 4(i)). 
Here $r_{\rm sub}(\theta)$ is defined as the radius where 
the hot dust temperature is $T_{\rm d}(\theta)=T_{\rm sub}=1500\,{\rm K}$ where $T_{\rm sub}$ is the silicate  grain sublimation temperature (e.g., 
\citealt{Ba87,LD93,Su06}). 
If this is the case, the sublimation radius could be elongated to extend to 
the central BH in the equatorial plane (see \citealt{KM10}), 
since the dust in the equatorial plane can survive because $T_{\rm d}
(\theta > \theta_{\rm torus}) < T_{\rm sub}$. 
Since such dust particles at $r_{\rm sub} (\theta > \theta_{\rm torus})$ 
can be heated to $\sim$1500 K due to illumination and emit NIR 
(see Fig. 4 (i)), the NIR luminosity is equal to the IR luminosity, i.e., 
$L_{\rm NIR,AGN}=L_{\rm IR,AGN}$ for all $\dot{m}_{\rm BH}$. Thus, we can 
just replace $L_{\rm IR,AGN}/L_{\rm bol, disc}$ assigned at the vertical axis 
in Fig. 2. with $L_{\rm NIR,AGN}/L_{\rm bol, disc}$. 

Second, we consider the case where the NIR emission comes from 
only a small part of the inner edge (Fig. 4(ii)). In this case, 
we define the NIR luminosity ($L_{\rm NIR, AGN}$) as the luminosity 
emitted by the hot dust ($T_{\rm d}=1000$--$1500\,{\rm K}$). The NIR 
luminosity is obtained as $L_{\rm NIR,AGN}=\left
(\frac{\cos{\theta_{\rm torus}-\cos{\theta_{\rm hot}}}}
{\cos{\theta_{\rm torus}}}\right)L_{\rm IR, AGN}$ where 
$\theta_{\rm hot}$ is defined as $T_{\rm d}(\theta_{\rm hot})
=1000\,{\rm K}$, as seen in Fig. 4(ii). The contribution of 
NIR emission decreases as the mass accretion rate is high, 
because for a large $\dot{m}_{\rm BH}$ the hot region at the inner 
edge of the torus (the red shaded region in Fig. 4(ii)) becomes 
smaller due to self-occultation of the super-Eddington accretion flow. 
As a result, we find that for a given $\theta_{\rm torus}$, 
the ratio $L_{\rm NIR,AGN}/L_{\rm IR,AGN}$ for super-Eddington AGNs 
with $\dot{m}_{\rm BH}\geq 10^{2}$ is much smaller than that 
for sub-Eddington AGNs with $\dot{m}_{\rm BH}=1$--$10$, as shown in 
Table 1.

Third, we consider the case where the dust temperature of the whole inner edge 
of the torus becomes lower than $T_{\rm sub}$ (Fig. 4(iii)). 
If this is the case, it is expected that there will be no NIR emission 
from the dusty torus for both sub- and super-Eddington AGNs. 
However, this prediction is in conflict with NIR observations of 
sub-Eddington AGNs (e.g., \citealt{El94,Gl06}). 
Thus, only for NIR-faint AGNs, the dust temperature at $r_{\rm in}$ 
might be below the dust sublimation temperature, because 
the inner radius is larger than the expected sublimation radius. 
To investigate this possibility, for NIR-faint AGNs, it is 
important to evaluate the time lag between the UV and NIR light 
by photometric monitoring observations (e.g., \citealt{Gl04,Mi04,Su06}). 

\begin{table}
\begin{center}
Table 1. Dependence of $L_{\rm NIR, AGN}/L_{\rm bol, disk}$ on 
$\theta_{\rm torus}$ and $\dot{m}_{\rm BH}$  \\[3mm]
\begin{tabular}{cccc}
\hline \hline
--- & $\theta_{\rm torus}=45^{\circ}$ & $60^{\circ}$ & $80^{\circ}$ \\
\hline 
$\dot{m}_{\rm BH}=1,10$ & 0.38 & 0.19 & 0.02 \\
$10^2$  & 0.08 & 0.01 & $8.4\times10^{-5}$ \\
$10^3$ & 0.028 & $8.3\times10^{-4}$ & $2.2\times10^{-8}$ \\
\hline
\end{tabular}
\end{center}
\end{table}

Taking into account the above results, we can conclude that a relatively low 
$L_{\rm NIR,AGN}/L_{\rm bol,disc}$, as well as $L_{\rm IR,AGN}/
L_{\rm bol,disc}$, is a sign of super-Eddington AGNs with 
$\dot{m}_{\rm BH}\geq 10^{2}$, unless the half opening angle of the torus 
is small ($\theta_{\rm torus} <65^{^\circ}$). 
We can also confirm that the predicted $L_{\rm NIR,AGN}/L_{\rm bol,disc}$ 
with $\dot{m}_{\rm BH}=1$--$10$ (Fig. 2, see also Table 1) almost coincides 
with the observed results of sub-Eddington quasars 
(e.g., \citealt{El94,Gl06}), assuming $L_{\rm bol,disc}=9L_{5100}$ 
where $L_{\rm 5100}$ is the AGN luminosity at $5100{\rm \AA}$ \citep{Ka00}. 

Finally, we discuss the evaporation of nearby dusty 
clouds irradiated by AGNs.
In the steady state for a dusty torus, cloud evaporation at the inner 
radius of the torus should balance the mass supply of the clouds from 
the outer region (see \citealt{KB88}). If the evaporation timescale, 
$t_{\rm evap}$, is longer than the orbital period of a cloud around SMBHs, 
$t_{\rm orb}$, the inner radius is determined by the sublimation radius, 
$r_{\rm sub}$. On the other hand, when $t_{\rm evap} \leq t_{\rm orb}$, 
the evaporation could be determined by the location of the inner edge of 
the torus. Thus, the inner radius of the dusty torus becomes 
larger than the sublimation radius. If this is the case, the NIR luminosity 
could be overestimated because the dust temperature is lower. To check this, 
we estimate the two timescales ($t_{\rm evap}$ and $t_{\rm orb}$) 
for the present model. Assuming $r_{\rm sub}=1.3L^{1/2}_{46}T^{-2.6}_{\rm sub}$, the orbital period of 
a cloud at $r_{\rm sub}$ is 
\begin{eqnarray}
t_{\rm orb}&=&2\pi r_{\rm sub}/v_{\phi} \nonumber \\
{}&=& 4.5\times 10^{4}L^{1/4}_{\rm 46}(L_{\rm bol,disc}/L_{\rm Edd})^{1/2}
\,{\rm yr}, 
\end{eqnarray}
where $v_{\phi}$ is the rotational velocity and $L_{46}=L_{\rm bol,disc}/
10^{46}\,{\rm erg}\,{\rm s}^{-1}$. 
Following \citet{PV95}, for a cloud of mass $M_{\rm c}=
10M_{\odot}$ and $T_{\rm dust}=1500\,{\rm K}$, the evaporation timescale is 
given by 
\begin{equation}
t_{\rm evap}=2.8\times 10^{5}(F(\theta_{\rm torus})/
10^{8}\,{\rm ergs}\,{\rm cm}^{-2}\,{\rm s}^{-1})^{-0.19}.  
\end{equation}
Note that $F(\theta_{\rm torus}=45^{\circ})$ at $r_{\rm sub}$ 
is $\approx 10^{8}\,{\rm ergs}\,{\rm cm}^{-2}\,{\rm s}^{-1}$ 
(eqs.(2) and (3)). 
Comparing these two timescales, we find $t_{\rm evap}\gg 
t_{\rm orb}$ for sub-Eddington AGNs with $\dot{m}_{\rm BH}
=1,10$(see \citealt{PV95} in details). 
On the other hand, for super-Eddington AGNs with $\dot{m}_{\rm BH}
=10^{3}$ and $L_{\rm bol,disc} > 10^{46}\,{\rm ergs}\,{\rm s}^{-1}$, 
the evaporation timescale is comparable to the orbital period of a cloud, 
i.e., $t_{\rm evap}\approx t_{\rm orb}$ (see eqs. (7) and (8)). If this is 
the case, evaporation of a dusty cloud would be fast enough to reduce 
the overall NIR emission. 
Therefore, AGNs with high $L_{\rm bol}$ and $\dot{m}_{\rm BH}$, 
such as bright narrow-line quasars, might have a much smaller ratio 
$L_{\rm NIR}/L_{\rm bol,disc}$, compared to the results of  Fig. 2. 

\section{Discussions}
\subsection{IR emission from NLR clouds}
So far, we have not taken into account the IR emission from 
narrow-line region (NLR) clouds. However, it is known that dust grains 
are present in the NLR of Seyferts (e.g., \citealt{DD88,NL93}). 
Thus, the dusty clouds in the NLR might be a source of IR emission 
and hence we must consider the contribution of IR emission from NLR clouds. 

NLR clouds are assumed to be distributed in a geometrically thin spherical 
shell with radius $r_{\rm NLR}$ in the range $0\leq \theta \leq 
\theta_{\rm torus}$. 
Assuming the covering factor of the NLR clouds, $f_{\rm NLR}$, and 
the optical depth of the NLR clouds, $\tau_{\rm NLR}$, 
the IR luminosity from NLR clouds can be given by 
\begin{eqnarray}
L_{\rm IR, NLR}&=&4\pi r^{2}_{\rm NLR}f_{\rm NLR}(1-e^{-\tau_{\rm NLR}})
\int_{0}^{\theta_{\rm torus}}F(\theta)\sin{\theta}d\theta.\nonumber \\
{}&=&f_{\rm NLR}(1-e^{-\tau_{\rm NLR}})
\left(1-\frac{L_{\rm IR,AGN}}{L_{\rm bol,disc}}\right),
\end{eqnarray}
where $L_{\rm IR,AGN}/L_{\rm bol,disc}$ is a function of 
$\theta_{\rm torus}$ (see eqs. (5) and (6)). 
In Fig. 4, we plot the ratio of the IR luminosity to the bolometric 
luminosity against $\theta_{\rm torus}$ for the NLR clouds, 
assuming $\tau_{\rm NLR}=1$ and $f_{\rm NLR}=0.1$. 
Note that the dust extinction of NLR emission is small (i.e., $\tau_{\rm NLR}
\leq 1$) for type 1 AGNs \citep{R00} and the typical
covering factor 
is about $10\%$ (e.g., \citealt{NL93,BL05}).

Figure 4 shows that the IR emission from the NLR clouds dominates above 
$\theta_{\rm torus}=75^{\circ}$ for $\dot{m}_{\rm BH}=1,10$, 
while for $\dot{m}_{\rm BH}=10^{2}$ and $\dot{m}_{\rm BH}=10^{3}$ 
the IR emission from NLR clouds is important above $\theta_{\rm torus}
=60^{\circ}$ and $\theta_{\rm torus}=50^{\circ}$, respectively.
This suggests that for super-Eddington AGNs with $\dot{m}_{\rm BH}\geq 10^{2}$, the IR emission from NLR clouds can dominate the IR emission
from the dusty torus. This is because the radiative flux of super-Eddington 
AGNs is highly collimated compared to sub-Eddington AGNs (see eqs. (2) and (3)). However, we should mention that the NLR size, $r_{\rm NLR}$, is 
about $10^{2}$ times as large as the sublimation radius 
(e.g., \citealt{Be02,Ne04,BL05}) and the dust temperature of 
NLR clouds is lower than the hottest dust temperature (i.e., 1500 K). 
Thus, the IR emission from NLR clouds would contribute mainly MIR 
(10--30 $\mu$m) emission (see \citealt{Sc08,Mor09}). 
In order to avoid contamination of the IR emission from NLR clouds, 
the ratio $L_{\rm NIR,AGN}/L_{\rm bol,disc}$ can be used to 
probe an inflated disk due to super-Eddington accretion, as mentioned 
in $\S 3.2$. 

\begin{figure}
\begin{center}
\includegraphics[height=6.5cm,clip]{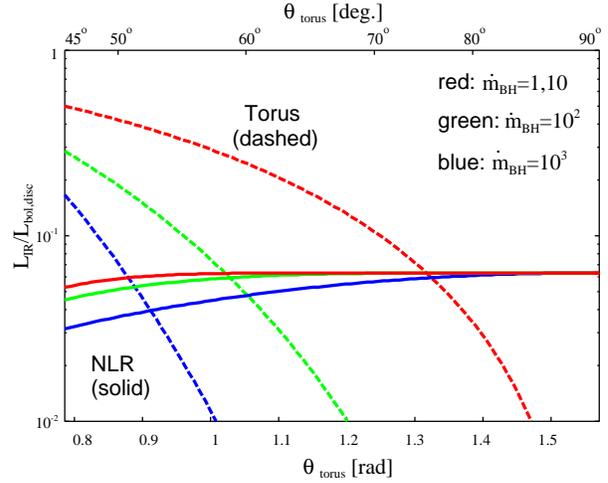}
\end{center}
\caption{
$L_{\rm IR,AGN}/L_{\rm bol,disc}$ against $\theta_{\rm torus}$ 
for various $\dot{m}_{\rm BH}$. 
The solid lines show the ratio of the IR luminosity and bolometric luminosity 
for the NLR clouds, assuming $\tau_{\rm NLR}=1$ and $f_{\rm NLR}=0.1$. 
The dashed lines show $L_{\rm IR}/L_{\rm bol,disc}$ 
for the dusty torus (see Fig. 2). 
}
\end{figure}

\subsection{Contribution of X-ray emission from the corona}
We here discuss the contribution of X-ray emission from the corona 
on the IR emission. For the model of X-ray emission from the corona, 
we adopt a simple lamp-post geometry model, where the X-ray source is 
located above the SMBH at some height $h_{\rm x}$
 (e.g., \citealt{GF91,NK01}). 
First, we examine the condition where the X-ray emission can reach the torus 
inner edge for super-Eddington accretion when $\theta_{\rm torus}\geq 
45^{\circ}$. 
If the light ray, which passes though the maximum thickness of the slim disc 
$h_{\rm slim}$ from the X-ray source located at $h_{\rm x}$, hits the torus 
inner edge, $r_{\rm sub}$, a large amount of X-ray emission is absorbed 
by the dust torus. Assuming $r_{\rm sub}\gg r_{\rm slim}$, this condition 
can be simply expressed by the following:
\begin{equation}
\frac{h_{\rm x}}{h_{\rm slim}} > 1-\tan(90^{\circ}-\theta_{\rm torus}), 
\end{equation}
where $r_{\rm slim}$ is the radius of the slim disc and 
$h_{\rm slim}=r_{\rm slim}$ because the half opening angle of 
the slim disc is $45^{\circ}$ (see $\S 2$). 
For example, the X-ray emission could produce the IR emission, 
if $h_{\rm x} > 0.4h_{\rm slim}$ and $h_{\rm x} > 0.8h_{\rm slim}$ 
for $\theta_{\rm torus}=60^{\circ}$ and $80^{\circ}$, respectively. 
According to the theory of slim discs, $r_{\rm slim}$ is about 
$10r_{\rm g}$ for $\dot{m}_{\rm BH}=10^{2}$. Note that for 
$\dot{m}_{\rm BH}=10^{3}$, $r_{\rm slim}$ is much larger than $10r_{\rm g}$ 
(e.g., \citealt{Wa00}). 
This may imply that the contribution of X-rays from the corona is 
not negligible if $h_{\rm x}$ is larger than $4$--$8r_{\rm g}$. 
The lower limit of $h_{\rm x}$ is comparable to $h_{\rm x}=6r_{\rm g}$, 
which is frequently used to discuss the iron-line reverberation in AGNs 
(e.g., \citealt{RB97}). 

We next investigate the contribution of X-rays from the corona 
on the IR emission when $h_{\rm x} > (4$--$8) r_{\rm g}$. 
The ratio of the radiation flux from the accretion disc, $F_{\rm disc}$, 
and the corona, $F_{\rm cor}$, at $\theta_{\rm torus}$ is expressed as 
\begin{equation}
\frac{F_{\rm disc}}{F_{\rm cor}}=\left(\frac{L_{\rm bol,disc}}
{L_{\rm x}}\right)\left(\frac{r_{\rm x}}{r_{\rm sub}}\right)^{2} \left \{
 \begin{array}{l}
2\cos{\theta_{\rm torus}}\,\,\, (\dot{m}_{\rm BH}=1,10), \\ \\
\frac{40\sqrt{2}}{7}\cos^{4}{\theta_{\rm torus}}\,\,\, (\dot{m}_{\rm BH}=10^{2}), \\ \\
\frac{160}{3}\cos^{9}{\theta_{\rm torus}}\,\,\, (\dot{m}_{\rm BH}=10^{3}). 
 \end{array}\right.
\end{equation}
Here $r_{\rm x}(\geq r_{\rm sub})$ is the distance from the X-ray source 
to the inner radius of the dusty torus. Assuming that the X-ray emission is 
isotropic, it is found that as $\theta_{\rm torus}$ and $\dot{m}_{\rm BH}$ 
increase, the X-ray emission from the corona begins to contribute to the
IR emission 
from the dusty torus. However, we should keep in mind that the dust opacity for 
X-ray radiation is lower than that for optical-UV radiation 
(e.g., \citealt{DL84,LD93}). Thus, the evaluated contribution of X-rays from 
the corona could be overestimated. 

\begin{figure}
\begin{center}
\includegraphics[height=5cm,clip]{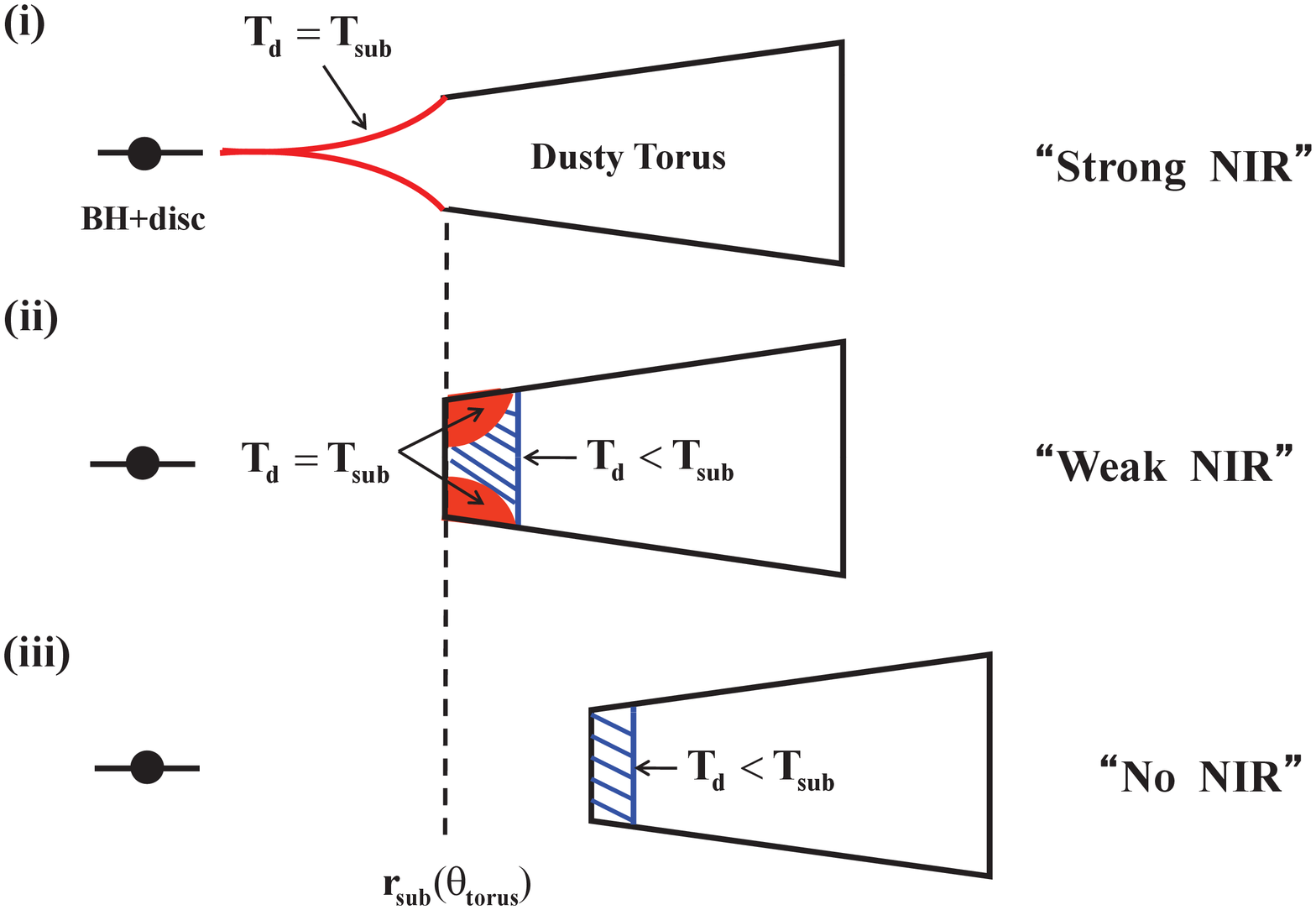}
\end{center}
\caption{
Three typical models for the shape of the inner edge of the torus. 
$T_{\rm d}$ and $T_{\rm sub}$ are the dust temperature and 
sublimation temperature, respectively. 
Note that $r_{\rm sub}(\theta_{\rm torus})$ is the sublimation 
radius at $\theta=\theta_{\rm torus}$. 
The red thick lines and shaded portions denote the region 
of NIR emission (i.e., $T_{\rm d}=T_{\rm sub}$). 
The blue shaded portions show the region of no NIR emission 
(i.e., $T_{\rm d} < T_{\rm sub}$).
}
\end{figure}

\subsection{Physical models of the dusty torus}
In $\S 3.2$, we considered three typical models for the configuration 
of the inner edge of the dusty torus (see Fig. 5). We here discuss how 
these three types can be understood physically. The following two physical 
quantities may be closely related to the shape of the inner part 
of the dusty torus. 
One is the sublimation radius, $r_{\rm sub}(\theta)$, which determines 
the region where hot dust can survive. In order to 
maintain a geometrically thick structure, energy input from stars 
and/or supernovae in the dusty torus is necessary 
(e.g., \citealt{WN02,Th05,KW08}). 
This energy input may occur outside of the critical radius, 
$r_{*}$, above which the torus is gravitationally unstable, 
if we adopt Toomre's stability criterion. 
Thus, the magnitude relation between $r_{\rm sub}(\theta_{\rm torus})$ 
and $r_{*}$ may determine the structure of 
the inner part of the dusty torus, as in the following three cases. 
(i) If $r_{*} \ll r_{\rm sub}(\theta_{\rm torus})$, the inner radius 
of the torus, $r_{\rm in}$, may be determined by $r_{\rm sub}(\theta)$. 
This corresponds to the case of Fig. 5(i). 
(ii) When $r_{*}\approx  r_{\rm sub}(\theta_{\rm torus})$, 
the structure of the dusty torus is like Fig. 5(ii).  
As mentioned above, the dust can survive at even $r < r_{\rm sub}
(\theta_{\rm torus})$, though the structure within $r_{*}$ is geometrically 
thin because there is no energy source in this region. Thus, the NIR emission 
from $r < r_{*}$ is negligible since the covering factor becomes small. 
That is, $r_{\rm in}$ is determined by the radius for which the dusty torus 
is gravitationally unstable. 
(iii) If $r_{*}  \gg r_{\rm sub}(\theta_{\rm torus})$, the dust 
temperature of the whole inner edge of the dusty torus becomes lower than 
$T_{\rm sub}$ because $r_{\rm in}=r_{*}$ is much larger than 
$r_{\rm sub}(\theta)$. If this is the case, the structure of 
the dusty torus is similar to Fig. 5(iii). 
By exploring the relation between the region of nuclear starbursts 
and that of the hot dust, we may find evidence revealing 
the formation of the dusty torus. 

\subsection{Evolutionary tracks in $L_{\rm IR,AGN}/L_{\rm bol,disc}
-\theta_{\rm torus}$ plane}
We now discuss the evolutionary tracks of SMBH growth in the 
$L_{\rm IR,AGN}/L_{\rm bol,disc}$--$\theta_{\rm torus}$ diagram. 
As mentioned in $\S 3.1$, although the formation mechanism of the obscuring 
torus is still a hotly debated issue (e.g., \citealt{KB88,PK92,OU99,OU01,WN02,WU05,
Wa09,Sc09}), \citet{WN02} proposed a dusty torus supported by turbulent 
pressure, in which the turbulence is produced by SN explosions (see also 
\citealt{KW08,Wa09}). In their model, the geometrical thickness of the torus, 
which is determined by the balance between the gravity of the central BH and the force due to turbulent pressure, is smaller for more massive BHs.
This tendency is consistent with observations in which the covering 
factor of the dusty torus decreases with increasing BH mass 
(e.g., \citealt{Ma07,No10}). 
This model also implies that $\theta_{\rm torus}$ increases with 
time as the BH grows. That is, the evolution proceeds from left to right 
in the $L_{\rm IR,AGN}/L_{\rm bol,disc}$--$\theta_{\rm torus}$ diagram (see Fig. 5). 
Here we consider two simple scenarios: 
(i) the super-Eddington growth dominated scenario 
where most of the mass of the SMBHs is supplied not through sub-Eddington 
accretion discs but through super-Eddington accretion discs. 
(ii) the sub-Eddington growth dominated scenario 
where the sub-Eddington accretion disc mainly feeds the SMBH. 
Figure 6 shows the evolutionary tracks for these two scenarios, 
assuming an initial opening angle of the torus of
$\theta_{\rm torus,init}=45^{\circ}$ and 
initially $\dot{m}_{\rm BH}=10^{3}$. 
We find from this figure 
that the AGNs stay in a super-Eddington phase ($\dot{m}_{\rm BH}>10^{2}$) 
for a long time (a wide range of $\theta_{\rm torus}$) in case (i), 
whereas in case (ii) 
the AGNs shift to a sub-Eddington phase 
when the torus is relatively thick.
The two distinct scenarios can be clearly understood 
from Fig. 3 as follows: 
for case (i) the evolution proceeds from top left $\to$ top right $\to$ 
bottom right, whereas for case (ii) it goes from top left $\to$ bottom left 
$\to$ bottom right. 
That is, the AGNs never undergo a phase of 
extremely low $L_{\rm IR,AGN}/L_{\rm bol,disc}$ for case (ii).
In contrast, 
in case (i), many AGNs 
can be identified as IR faint objects.
The NIR luminosity of such AGNs is also very small.
The dispersion of $L_{\rm IR,AGN}/L_{\rm bol,disc}$ 
as well as $L_{\rm NIR,AGN}/L_{\rm bol,disc}$ 
for case (i) is significantly larger than that for case (ii).
If the majority of high-$z$ quasars ($z>7$)
are NIR-faint, super-Eddington growth may be inevitable 
for SMBH growth in the early Universe. 
In order to compare the predictions with observations in detail, we must elucidate the 
evolution of $L_{\rm NIR,AGN}/L_{\rm bol,disc}$ based on the coevolution 
model of SMBH growth and a circumnuclear disc such as \citet{KW08}. 
This is a subject for future work. 

\begin{figure}
\begin{center}
\includegraphics[height=6.5cm,clip]{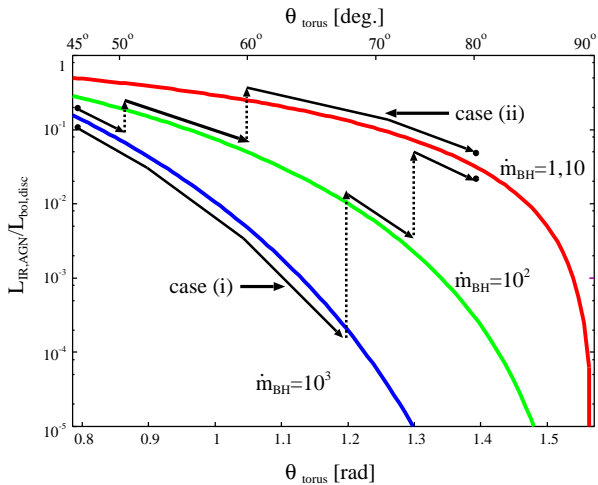}
\end{center}
\caption{
Schematic diagram of two evolutionary scenarios, 
super-Eddington growth dominated scenario (case (i)) 
and sub-Eddington growth dominated scenario (case(ii)). 
The evolution proceeds from left to right, i.e., as the mass of the BH 
increases, $\theta_{\rm torus}$ increases (see text). 
We assume that the initial opening angle of the torus is 
$\theta=45^{\circ}$ and $\dot{m}_{\rm BH}=10^{3}$.
}
\end{figure}

\subsection{Comparison with observations}
In this paper, we predicted that super-Eddington AGNs with 
$\dot{m}_{\rm BH}\geq 10^{2}$ have relatively low $L_{\rm NIR,AGN}
/L_{\rm bol,disc}$. 
So far, more than 30 quasars have been discovered at $z\approx 6$ 
(e.g., \citealt{Fa01,Fa06,Go06,Wi07,Wi10}). 
\citet{Ji10} found that two are unusually NIR-faint quasars, 
i.e., their ratio of $L_{\rm NIR}/L_{\rm 5100}$ is one order of magnitude 
smaller than the average for ordinary quasars, although their properties 
are similar to those of low-$z$ quasars in the rest-frame ultraviolet/optical 
and X-ray bands (e.g., \citealt{Fa04,Ji06,Sh06}). 
More interestingly, the two quasars have the smallest BH mass ($M_{\rm BH}
\approx 10^{8}M_{\odot}$) and the highest Eddington ratios ($L_{\rm bol,AGN}
/L_{\rm Edd}\sim 2$) of $z\sim 6$ quasar samples. 
These features can be nicely explained by our interpretation, i.e., 
the faint NIR emission for high Eddington AGNs. 
We should note that since NIR-faint quasars can be explained by sub-Eddington 
AGNs with extremely thin tori (i.e., $\theta_{\rm torus}>85^{\circ}$), 
we need to measure the thickness of a dusty torus to confirm whether 
NIR-faint quasars are really super-Eddington AGNs ($\dot{m}_{\rm BH}
\geq 10^{2}$). 
Based on the kinematics of cold molecular gas in a torus, 
it would be possible to evaluate the thickness of a torus by ALMA, 
because the ratio of velocity dispersion and rotational 
velocity indicates the scale height of the torus (see \citealt{WN02,WT05}). 
On the other hand, \citet{Ji10} interpreted these two objects as dust-free 
quasars. However, this seems to contradict the observations, which suggests 
that NIR-faint quasars possess super-metallicity as do other $z\sim 6$ quasars and high-$z$ AGNs (e.g., \citealt{Pe02,Fr03,Ju09,HF93,Na06a,Na06b,Ma09}). 
Also, \citet{Ha10} recently reported quasars with unusually faint NIR 
emission at $z=1.5$--$3$, when the universe was 2--4 Gyr old. Thus, it seems 
likely that NIR-faint quasars are exclusively dust-free AGNs. 
As an alternative scenario, a large fraction of gas in a dusty torus would be 
ejected from host galaxies as a result of strong radiation pressure 
from the brightest AGNs, such as quasars \citep{OU99,OU01,WU05}. 
In this case, NIR-faint quasars may be explained as being quasars 
without dusty tori around the SMBH. This possibility could be confirmed 
by investigating the presence of dusty tori in NIR-faint quasars 
using ALMA. 

In the local universe, NLS1s are thought to be AGNs of a high 
$L_{\rm bol, disc}/L_{\rm Edd}$ system. 
One NLS1 (Ark 564) shows no NIR emission \citep{RM06},
which is consistent with our predictions. 
However, most NLS1s tend to have NIR emission similar to or 
stronger than that of ordinary Seyfert galaxies (e.g. \citealt{Ry07}). 
One of the reasons for the discrepancy between our predictions 
and the observations is that NLS1s with $\dot{m}_{\rm BH}\geq 10^{2}$ may be 
absent (or their fraction is very small). However, some NLS1s have 
a high mass accretion rate with $\dot{m}_{\rm BH}\geq 10^{2}$ 
(e.g., \citealt{CK04,Ha08}). Another possibility is that these discrepancies 
may suggest misalignment between the accretion discs and the 
dusty torus, as mentioned in $\S 3.1$. 
Lastly, we note the possibility of contamination of hosts and circumnuclear 
starbursts in the NIR band, if the host luminosity is comparable to or exceeds
the AGN luminosity, as in Seyfert galaxies. Although it may be hard to find 
many NLS1s with low $L_{\rm NIR,AGN}/L_{\rm bol,disc}$ in current observations, in future studies it is worth examining this issue by NIR (rest frame) 
observations with high spatial resolution.

\section{Summary}
In this work, we propose a new method to search for candidate 
galaxies in which super-Eddington growth occurs. 
In particular, taking account of the dependence of 
the polar angle on the radiation flux of accretion flows, 
we investigate the properties of reprocessed IR emission (as well as 
NIR emission) from the inner edge of the dusty torus. As a result, 
we find that a relatively low ratio of AGN IR (as well as NIR) 
luminosity and disc luminosity is a genuine property 
that can be used for exploring for candidates of super-Eddington AGNs with 
$\dot{m}_{\rm BH}\geq 10^{2}$. This method 
is especially powerful for searching for high-$z$ super-Eddington AGNs 
because the direct measurement of $\dot{m}_{\rm BH}$ is 
difficult for high-$z$ AGNs compared with nearby AGNs.

The following are the main results of the present paper: 
\begin{enumerate} 
\item We find that the ratio of the AGN IR luminosity, $L_{\rm IR, AGN}$, 
and the disc bolometric luminosity, $L_{\rm bol, disc}$, decreases as 
the half opening angle of the dusty torus, $\theta_{\rm torus}$, 
increases and the mass accretion rate normalized by the Eddington rate, 
$\dot{m}_{\rm BH}$, increases. The dependence of the ratio on $\dot{m}_{\rm BH}$ 
is caused by the self-occultation effect, which attenuates the illumination 
of the torus via the self-absorption of the emission from the inner disc 
surface at the outer region of super-Eddington accretion discs.
The ratio in super-Eddington AGNs with $\dot{m}_{\rm BH}\geq 10^{2}$
is more than one order of magnitude lower than 
that in sub-Eddington AGNs with $\dot{m}_{\rm BH}=1$--$10$
for $\theta_{\rm torus} > 65^{\circ}$ ($\dot{m}_{\rm BH} = 10^2$)
and for $\theta_{\rm torus} > 55^{\circ}$ ($\dot{m}_{\rm BH} = 10^3$).
A small value of $L_{\rm IR, AGN}/L_{\rm bol, disc}$ 
could be a signature of super-Eddington accretion with 
$\dot{m}_{\rm BH}\geq 10^{2}$, since it cannot be realized in 
sub-Eddington AGNs, unless the torus has quite a wide opening angle,
$\theta_{\rm torus}>85^\circ$. 

\item We also examined the properties of the AGN near-IR luminosity 
($L_{\rm NIR,AGN}$) radiated from hot dust at $>1000\,{\rm K}$, 
because it may be hard to detect pure AGN IR emission from the inner 
edge of a torus because of contamination of the nuclear starburst. 
In this case, $L_{\rm NIR, AGN}/L_{\rm bol,disc}$ of super-Eddington AGNs with $\dot{m}_{\rm BH}\geq 10^{2}$ 
is at least one order of magnitude smaller than that for sub-Eddington AGNs over a wide range of 
half opening angle ($\theta_{\rm torus}\geq 65^{\circ}$). 
This is true regardless of which of the three typical models of the dusty torus is used. 
\end{enumerate}

\section*{Acknowledgments} 
We appreciate the useful comments of the anonymous referees that reviewed this paper. 
We are grateful to M. Umemura and T. Nagao for useful comments and 
discussions. This work is supported in part by the Ministry of Education, 
Culture, Sports, Science, and Technology (MEXT) Research Activity 
Start-up 2284007 (NK) and Young Scientist (B) 20740115 (KO).

\bsp

\label{lastpage}

\end{document}